\documentclass[aps,prl,floatfix,twocolumn,nopacs,superscriptaddress]{revtex4-1}

\usepackage[dvipdfmx]{color}

\usepackage{graphicx}
\usepackage{epstopdf}

\usepackage{amsfonts,amssymb,amsmath}

\usepackage{hyperref}

\begin{document}

\title{
Current scaling of the topological quantum phase transition \\
between a quantum anomalous Hall insulator and a trivial insulator}

\author{Minoru~Kawamura}
\email[]{minoru@riken.jp}
\affiliation{
	RIKEN Center for Emergent Matter Science (CEMS), Wako  351-0198, Japan\\
}

\author{Masataka~Mogi}
\affiliation{
	Department of Applied Physics 
	and Quantum-Phase Electronics Center (QPEC), University of Tokyo,  Tokyo 113-8656, Japan\\
}

\author{Ryutaro~Yoshimi}
\affiliation{
	RIKEN Center for Emergent Matter Science (CEMS), Wako  351-0198, Japan\\
}

\author{Atsushi~Tsukazaki}
\affiliation{
	Institute for Materials Research, Tohoku University, Sendai 980-8577, Japan\\
}

\author{Yusuke~Kozuka}
\affiliation{
	Department of Applied Physics 
	and Quantum-Phase Electronics Center (QPEC), University of Tokyo,  Tokyo 113-8656, Japan\\
}

\author{Kei~S.~Takahashi}
\affiliation{
	RIKEN Center for Emergent Matter Science (CEMS), Wako  351-0198, Japan\\
}
\affiliation{
	PRESTO, Japan Science and Technology Agency (JST),  Tokyo 102-0075, Japan
}

\author{Masashi~Kawasaki}
\affiliation{
	RIKEN Center for Emergent Matter Science (CEMS), Wako  351-0198, Japan\\
}
\affiliation{
	Department of Applied Physics 
	and Quantum-Phase Electronics Center (QPEC), University of Tokyo,  Tokyo 113-8656, Japan\\
}

\author{Yoshinori~Tokura}
\affiliation{
	RIKEN Center for Emergent Matter Science (CEMS), Wako  351-0198, Japan\\
}
\affiliation{
	Department of Applied Physics 
	and Quantum-Phase Electronics Center (QPEC), University of Tokyo,  Tokyo 113-8656, Japan\\
}
\affiliation{
	Tokyo College, University of Tokyo, Tokyo 113-8656, Japan \\
}

\date{\today}

\begin{abstract}
We report a current scaling study of a quantum phase transition 
between a quantum anomalous Hall insulator and a trivial insulator
 on the surface of a heterostructure film of magnetic topological insulators.
The transition was observed by tilting the magnetization
while measuring the Hall conductivity $\sigma_{xy}$.
The transition curves of $\sigma_{xy}$ 
taken under various excitation currents cross each other at a single point,
exemplifying a quantum critical behavior of the transition.
The slopes of the transition curves follow a power law  dependence
of the excitation current, giving a scaling exponent.
Combining with the result of the previous temperature scaling study,
critical exponents $\nu$ for the localization length and $p$
for the coherence length are separately evaluated
as $\nu$ = 2.8 $\pm$ 0.3 and $p$ = 3.3 $\pm$ 0.3.
\end{abstract}

\maketitle

Topological phases of matter have been one of the central research topics
in contemporary condensed-matter physics\cite{MZHasan_RMP2010, XLQi_RMP2011, BYan_AnnuRev2017}.
A quantum phase transition (QPT) between two topologically distinct phases
can be characterized by a change in the topological index of gapped bulk energy bands
even if the symmetry does not change.
According to the bulk-edge correspondence,
the QPT is accompanied by a closing of the energy gap at the quantum critical point (QCP),
resulting in abrupt changes in physical quantities such as transport coefficients.

Quantum anomalous Hall (QAH) insulator is one of the most distinct topological phases
materialized on the surface
of ferromagnetic topological insulators\cite{RYu_Science2010, CZChang_Science2013, JGCheckelsky_NPhys2014, XKou_PRL2014, CZChang_NMat2015, MMogi_APL2015, SGrauer_PRL2017, YTokura_NRevPhys2019}.
Magnetic exchange interaction between itinerant surface electrons 
and localized magnetic moments opens an energy gap in the dispersion relation
of the surface state, stabilizing the QAH insulator phase.
The QAH insulator possesses a one-dimensional chiral edge channel
running on the side surface of a sample
which contributes to the quantized Hall conductivity of $e^2/h$,
where $e$ is the elementary charge and $h$ is the Planck's constant.
The QAH effect has been studied intensively
in a family of magnetically-doped topological insulators
Cr- and/or V-doped (Bi,Sb)$_2$Te$_3$ and their heterostructure films.
Recently, materials hosting the QAH effect has been expanded to include
a stoichiometric magnetic topological insulator MnBi$_2$Te$_4$\cite{CLiu_NMater2020, YDeng_Science2020} 
and twisted bilayer graphene\cite{MSerlin_Science2020, ALSharpe_Science2019}.
Recent experimental studies have explored various insulator phases
in  magnetic topological insulators
such as an axion insulator under anti-parallel magnetizations\cite{MMogi_NMater2017, MMogi_SciAdv2017},
a trivial insulator stabilized by the hybridization of top and bottom 
surface states\cite{YZhang_NPhys2010, MKawamura_PRB2018, XKou_NCommun2015},
and an Anderson insulator\cite{CZChang_PRL2016}.
Since then, the QPTs between the QAH insulator and the other insulator phases
have attracted considerable attentions\cite{YFeng_PRL2015, MKawamura_PRB2018, CZChang_PRL2016, XKou_NCommun2015, CLiu_NMater2020}.

The QPTs between the QAH insulator and the other insulator phases 
are signaled by changes in the Hall conductivity $\sigma_{xy}$
which is a $e^2/h$ multiple of the Chern number\cite{DJThouless_PRL1982}.
The QPTs possess many similarities to the plateau-plateau transitions of 
the quantum Hall (QH) effect in semiconductor
two-dimensional systems\cite{HAoki_PRL1985, AMMPruisken_PRL1988,
SATrugman_PRB1983, DHLee_PRL1993,
JTChalker_1988, KSlevin_PRB2009,
BHuckestein_RMP1995, SLSondhi_RMP1997, 
HPWei_PRL1988, SKoch_PRL1991, MHilke_Nature1998, WLi_PRL2005,
DGPolyakov_PRB1993, HPWei_PRB1994, WPan_PRB1997,
DoDoo-Amoo_2014}.
The QH transitions are characterized
by a divergence of the localization length $\xi \propto |E - E_{\rm c}|^{-\nu}$
as the QCP is approached,
where $E_{\rm c}$ is a critical energy and $\nu$ is a critical exponent
for the localization length.
Theoretical studies have shown the universal exponent $\nu$ = 2.6
for a quantum network of chiral edge channels while $\nu$ = 4/3 for a classical
network\cite{SATrugman_PRB1983, DHLee_PRL1993, 
BHuckestein_RMP1995, SLSondhi_RMP1997, JTChalker_1988, KSlevin_PRB2009}.
Because a critical exponent of a QPT is believed to reflect
only fundamental symmetries and dimensionality of a system 
regardless of its detailed physical realization, 
the critical exponent is predicted to be $\nu$ = 2.6 \cite{JWang_PRB2014}
for the QPT between the QAH insulator and an axion insulator,
which is the same as the result of the Chalker-Coddington model
for the QH transitions\cite{JTChalker_1988, KSlevin_PRB2009}.
A Berezinskii-Kosterlitz-Thouless (BKT) type phase transition
and a slightly different critical exponent $\nu$ = 2.4 
are also theoretically proposed when random magnetic domains are involved\cite{CZChen_PRL2019}.

Experimentally, the QPTs have been
studied in ferromagnetic topological insulator thin films
by reversing the magnetization\cite{XKou_NCommun2015}, 
by changing the carrier density\cite{CZChang_PRL2016}, 
or by changing the magnetization direction\cite{MKawamura_PRB2018}.
The critical behaviors of the QPTs were studied by the temperature ($T$) scaling
of the transport coefficients where the sharpness of the transition ($\Delta E^{-1}$)
scales as $\Delta E^{-1} \sim T^{-\kappa}$ with a critical exponent $\kappa$.
In these studies, $\kappa$ ranging from 0.22 to 0.62 are reported.
However, the exponent $\kappa$ 
is a combination value ($= p/2\nu$, as shown below) of $\nu$ and another exponent $p$
for the coherence length ($l_\phi \sim T^{-p/2}$) \cite{HAoki_PRL1985}.
To compare the experimentally observed critical exponent with the theories,
an independent measurement of $p$ is essential.

In this paper, the two critical exponents $p$ and $\nu$ of the QPT 
between the QAH insulator and the trivial insulator
 are separately evaluated employing a current scaling technique developed
in the earlier studies of the QH transitions\cite{DGPolyakov_PRB1993, HPWei_PRB1994, WPan_PRB1997, DoDoo-Amoo_2014}.
The QPT was driven by tilting the magnetization in a heterostructure film
of magnetic topological insulator as similar to a previous study\cite{MKawamura_PRB2018}.
Because only the magnetization component perpendicular to the film plane contribute
to the exchange gap energy, the exchange gap energy can be tuned by tilting the magnetization
with an assistance of external magnetic fields.
The QPT occurs when the exchange gap energy meets the hybridization energy caused
by the coupling of the top and bottom surface states\cite{MKawamura_PRB2018, JWang_PhysScr2015}.
In this method, all the magnetic moments are forced to align in the direction of the  
external field, therefore complications arising from the magnetic domain formation are avoided. 
We observe transitions of $\sigma_{xy}$  from $e^2/h$ to zero
as the magnetization is tilted away from the direction perpendicular to the film plane.
By analyzing the current dependence of the transition curves,
a current scaling exponent is determined.
Combining with the result of the previous temperature scaling study,
the critical exponents $\nu$ and $p$ are separately evaluated as $\nu$ = 2.8 $\pm$ 0.3
and $p$ = 3.3 $\pm$ 0.3.

In the theoretical studies\cite{BHuckestein_RMP1995, KSlevin_PRB2009},
the exponent $\nu$ is evaluated numerically using the finite-size scaling method
changing the system size.
In the experimental studies, instead of the system size,
the phase coherence length $l_\phi$ is changed
through temperature $T$.
In a diffusive system, the phase coherence time $\tau_{\phi}$  follows a power law
as $\tau_\phi \sim T^{-p}$, leading to $l_\phi \sim T^{-p/2}$.
Then, by putting $l_\phi$ as a cutoff length, 
the sharpness of the transition follows a power law
$\Delta E^{-1} \propto T^{-\kappa}$ 
with $\kappa = p/2\nu$\cite{HAoki_PRL1985, HPWei_PRL1988}.
In the current scaling analysis, the mean energy drop between coherent regions
is regarded as an effective temperature: $k_{\rm B}T_{\rm eff} = eRI/(L/l_\phi)$,
where $k_{\rm B}$ is the Boltzmann's constant, 
$R$ the sample resistance,  $I$ the excitation current, and $L$ the sample size.
Then,  $l_\phi$ is related to $I$ as  $l_\phi \propto I^{-p/(p+2)}$.
Consequently,  the sharpness of the transition follows a power law
 $\Delta E^{-1} \propto I^{-b}$
 with $b = p/(p+2)\nu$.
Therefore, by measuring $\kappa$ and $b$ experimentally,
$\nu$ and $p$ can be determined separately\cite{DGPolyakov_PRB1993, HPWei_PRB1994}.

\begin{figure}[t]
	\includegraphics[width=.95\columnwidth]{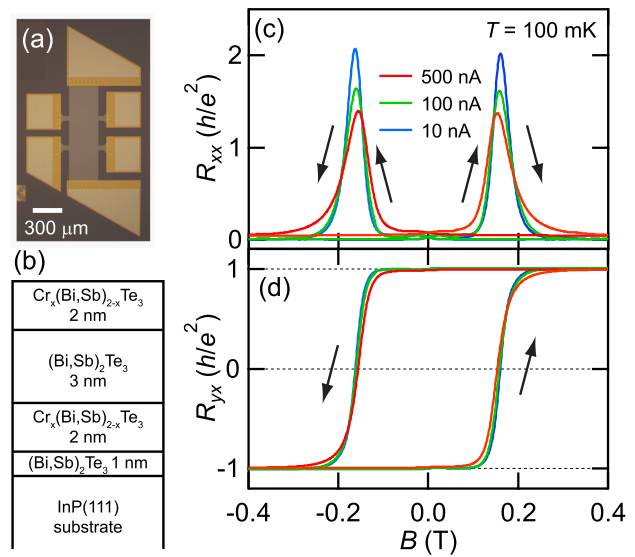}
	\caption{(a) Photograph of the Hall bar sample.
			(b) Schematic structure 
			of the magnetic topological insulator heterostructure film.
			The Cr content $x$ = 0.24.
			(c)(d) Magnetic field dependence of $R_{xx}$ (c) and $R_{yx}$ (d) at $T$ = 100 mK
			measured using the excitation current $I$ = 10 nA, 100 nA, and 500 nA.
			The data were taken by sweeping the field in the positive and negative directions
			indicated by arrows 	at a rate of 0.03 T/min.
			}
	\label{fig1}
\end{figure}

Experiments were conducted using
an identical  Hall-bar sample [Fig.~\ref{fig1}(a)] to the previous study\cite{MKawamura_PRB2018}.
The Hall bar was made from a heterostructured film of 
 (Bi,Sb)$_2$Te$_3$ sandwiched by 2-nm-thick Cr-doped (Bi,Sb)$_2$Te$_3$
grown on InP(111) substrate by molecular beam epitaxy (MBE)
as schematically shown in Fig.~\ref{fig1}(b).
The total thickness of the film was 8 nm.
Bi/Sb ratio was adjusted so that the Fermi energy lies close to the charge neutrality point.
Details of the sample preparation is described elsewhere\cite{MMogi_NMater2017}.
Low temperature transport measurements were conducted at
 $T$ = 100 mK using a dilution refrigerator
equipped with a single-axis sample rotator.
A low-frequency lock-in technique was employed for the resistance measurement.
The excitation current was applied through a resistance of 10 M$\Omega$ or 1 M$\Omega$
connected to the sample in series.

Figures~\ref{fig1}(c) and ~\ref{fig1}(d) respectively show the magnetic field dependence
of $R_{xx}$ and $R_{yx}$ under application of excitation current 
$I$ = 10 nA, 100 nA, and 500 nA.
The magnetic field was applied almost perpendicular to the film plane.
The QAH effect with the quantized Hall resistance $h/e^2$ is clearly observed.
The sign of the Hall resistance changes when the magnetization is reversed,
accompanied by a peak in $R_{xx}$.
As the excitation current is increased, the $R_{xx}$ peaks are broadened
and the slopes of the Hall resistance curves become gentle at the magnetic fields
near the coercive fields.
At $I$ = 500 nA, $R_{xx}$ is  lifted off from zero at $B$ = 0.5 T 
and $R_{yx}$ is slightly decreased from $h/e^2$.
Thus, the increase in excitation current gives qualitatively similar effects
on the $R_{xx}$-$B$ and $R_{yx}$-$B$ curves as
the increase in temperature.

\begin{figure}[t]
	\includegraphics[width=.95\columnwidth]{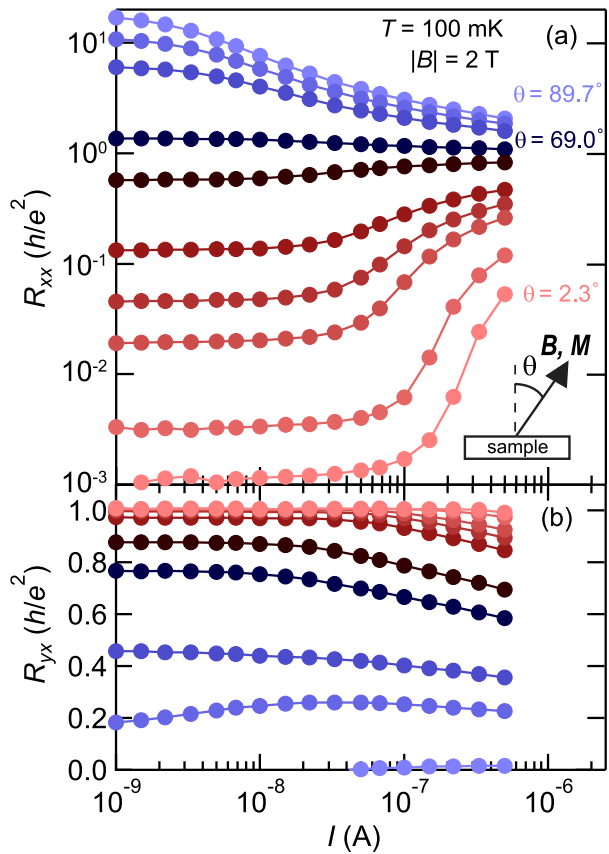}
	\caption{
			(a)(b) Current dependence of $R_{xx}$ (a) and $R_{yx}$(b) at $|B|$ = 2 T
			and $T$ = 100 mK under various magnetic field angles $\theta$.
			Data for $\theta$ = 2.3$^\circ$, 28.2$^\circ$,
			42.7$^\circ$, 47.9$^\circ$, 53.2$^\circ$, 63.7$^\circ$, 69.0$^\circ$,
			78.1$^\circ$, 82.5$^\circ$, and 89.7$^\circ$ are plotted from bottom to top
			in (a).
			$\theta$ is measured from the direction perpendicular
			to the film as depicted in the inset in (a).
			}
	\label{fig2}
\end{figure}

Figures~\ref{fig2}(a) and ~\ref{fig2}(b) show the current dependence 
of $R_{xx}$ and $R_{yx}$ under various magnetization angles, respectively.
In this measurement,
an external magnetic field $|B|$ = 2 T,
which is much larger than the coercive field of the present sample, was applied
and was rotated to tilt the magnetization direction using the single-axis sample rotator.
The tilted angle $\theta$ is measured from the direction perpendicular to the film plane. 
When the magnetization is almost perpendicular to the film plane ($\theta$ = 2.3$^\circ$),
the value of $R_{xx}$ increases steeply as $I$ exceeds 100 nA,
accompanied by a deviation of $R_{yx}$ from $h/e^2$.
This corresponds to the current-induced breakdown
of the QAH effect\cite{MKawamura_PRL2017}.
When the magnetization is almost parallel to the film plane ($\theta$ = 89.7$^\circ$),
the value of $R_{xx}$ decreases with increasing the excitation current.
This current dependence is a typical behavior of a trivial insulator.
At the intermediate angles  around $\theta$ = 69.0$^\circ$
where $R_{xx}$ is close to $h/e^2$,
$R_{xx}$ is nearly independent of $I$.

\begin{figure}[t]
	\includegraphics[width=.95\columnwidth]{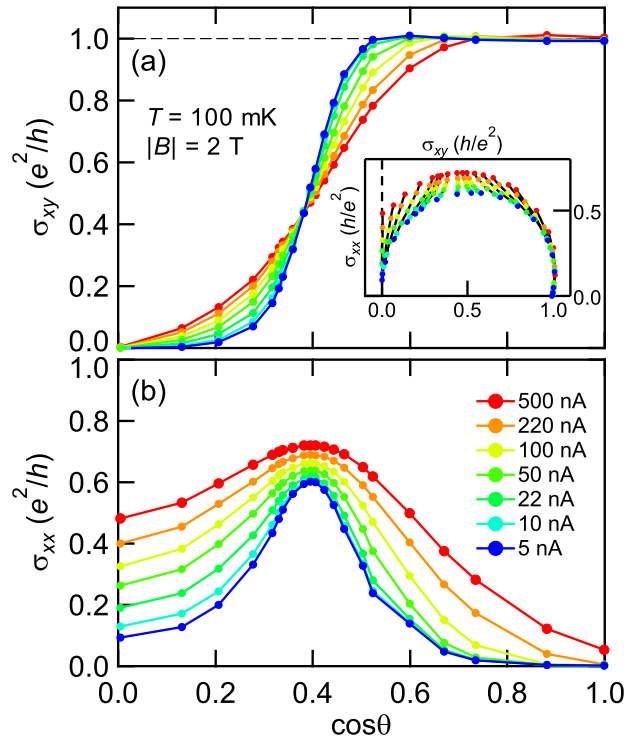}
	\caption{
			(a)(b) Magnetization angle dependence of $\sigma_{xy}$ (a) and $\sigma_{xx}$ (b)
			plotted as a function of $\cos\theta$.
			The conductivity tensor components are converted from the resistance data
			shown in Fig.~\ref{fig2} using the relation 
			$\sigma_{xy} = \rho_{yx}/(\rho_{xx}^2 + \rho_{yx}^2)$ and
			$\sigma_{xx} = \rho_{xx}/(\rho_{xx}^2 + \rho_{yx}^2)$.
			The data for several excitation currents ranging from 5 nA to 500 nA are shown.
			The inset in (a) shows a parametric plot of ($\sigma_{xy}$, $\sigma_{xx}$).
			Each black curve connects the points belonging to the same magnetization angle.
			}
	\label{fig3}
\end{figure}

To see the critical behavior of the QPT,
the conductivity tensor components $\sigma_{xx}$ and $\sigma_{xy}$  calculated
from $R_{xx}$ and $R_{yx}$  are plotted as a function of $\cos\theta$ in Fig.~\ref{fig3}.
The Hall conductivity $\sigma_{xy}$ transits from $e^2/h$ to zero 
as $\cos\theta$ is decreased [Fig.~\ref{fig3}(a)].
The transition in $\sigma_{xy}$ is accompanied by a peak
in $\sigma_{xx}$ [Fig.~\ref{fig3}(b)], reflecting the energy gap closing at the QCP.
The QPT becomes sharp as the current is decreased.
Several $\sigma_{xy}$-$\cos\theta$ curves for various $I$ cross almost at a single point,
exemplifying the QPT between the QAH insulator and the trivial insulator.
The crossing point is the QCP of the transition.
The crossing point ($\cos\theta$, $\sigma_{xy}$) = (0.38, 0.43~$e^2/h$)
is almost the same as the QCP in the previous temperature dependence 
measurement\cite{MKawamura_PRB2018}.
The parametric plot of ($\sigma_{xy}$, $\sigma_{xx}$) for various $I$
in the inset of Fig.~\ref{fig3}(a) shows a flow as similar to the temperature scaling flow\cite{JGCheckelsky_NPhys2014, AMMPruisken_PRL1988, MKawamura_PRB2018}:
 ($\sigma_{xx}$, $\sigma_{xy}$) tends to converge either (0, 0) or ($e^2/h$, 0)
 with decreasing current with an unstable point at around (0.5~$e^2/h$, 0.5~$e^2/h$).

\begin{figure}[t]
	\includegraphics[width=.95\columnwidth]{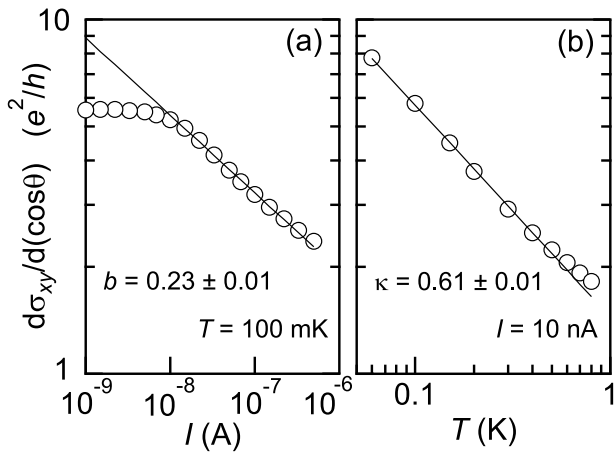}
	\caption{
			(a)(b) 
			The slope of the  $\sigma_{xy}$-$\cos\theta$ curve
			at the QCP $d\sigma_{xy}/d\cos\theta$
			plotted as a function of excitation current (a)
			and temperature (b).
			The data in (b) are taken from Ref. \onlinecite{MKawamura_PRB2018}.
			The solid lines are the fitting results to
			$d\sigma_{xy}/d\cos\theta \propto I^{-b}$ in (a)
			 and $\propto T^{-\kappa}$ in (b), respectively.
			The measurement temperature was $T$ = 100 mK in (a).
			The excitation current was $I$ = 10 nA in (b). 
			}
	\label{fig4}
\end{figure}

Next, we analyze how the sharpness of the QPT changes with the excitation current.
Figure~\ref{fig4}(a) shows the $I$ dependence
of the slope $d \sigma_{xy}/d\cos\theta (I)$ at the QCP.
For comparison, the temperature dependence of $d \sigma_{xy}/d\cos\theta (T)$
taken from the previous study\cite{MKawamura_PRB2018} is shown in Fig.~\ref{fig4}(b).
The slope $d \sigma_{xy}/d\cos\theta (I) $ increases  with decreasing $I$
and turns to saturate below $I \sim 10$ nA as shown in Fig.~\ref{fig4}(a). 
Because the current dependence measurement was conducted at $T$ = 100 mK, 
the saturated value shows a good agreement with
$d \sigma_{xy}/d\cos\theta$ ($T$ = 100 mK) in Fig.~\ref{fig4}(b). 
The saturation in the $I$ dependence indicates that
$d \sigma_{xy}/d\cos\theta (I) $ is limited by the thermal excitation in the range $I < 10$ nA
while $\sigma_{xy}/d\cos\theta (I) $ is dominated by  the excitation current
 in the range $I \ge$ 10 nA.
For $I \ge $ 10 nA, the slope follows a power law current dependence
$d \sigma_{xy}/d\cos\theta (I) \sim I^{-b}$ 
with a current scaling exponent $b$ = 0.23 $\pm$ 0.01.
Combining $b$ with the temperature scaling exponent $\kappa$ = 0.61 $\pm$ 0.01\cite{MKawamura_PRB2018},
the critical exponents $\nu$ for the localization length
and $p$ for the coherence length
are separately yielded as $\nu$ = 2.8 $\pm$ 0.3 and $p$ = 3.3 $\pm$ 0.3.

The value $\nu$  = 2.8 $\pm$ 0.3 
is reasonably close to the result of the Chalker-Coddington model
$\nu$ = 2.6 \cite{JTChalker_1988, KSlevin_PRB2009}
and those reported in the QH transitions in InGaAs/InP heterostructure\cite{HPWei_PRB1994, WPan_PRB1997} but is apparently larger 
than the exponent $4/3$ for the classical percolation
in two-dimensional systems \cite{SATrugman_PRB1983, DHLee_PRL1993}.
This result indicates that the transition between the QAH insulator
and the trivial insulator can be modeled by a quantum percolation
of a network of chiral edge channels where quantum tunneling at the saddle points
dominates the critical behavior\cite{JTChalker_1988, JWang_PRB2014}.
According to the earlier studies of the QH transitions\cite{DoDoo-Amoo_2014, WLi_PRB2010},
$l_\phi > \xi$ is an essential condition to obtain the universal critical exponent.
A rather short localization length due to strong disorder in the surface state of
the magnetic topological insulators\cite{Lee_PNAS2015, MKawamura_PRL2017}
probably assists this condition to be satisfied.
Note that there is no magnetic multi-domain structure at the QCP in the present experiment due to the externally applied magnetic field.
Therefore, the BKT-type phase transition discussed in Ref. \onlinecite{CZChen_PRL2019} is not likely to occur.

The value $p$ = 3.3 $\pm$ 0.3 is considerably large compared to 
 $p \sim$ 2 reported in the earlier studies of QH transitions 
in the  InP/InGaAs heterostructures\cite{HPWei_PRB1994, WPan_PRB1997}.
A recent study\cite{DoDoo-Amoo_2014} of the QH transitions in GaAs/AlGaAs heterostructures
reports a large variation of $p$ from 0.5 to 3.9 with decreasing mobility.
The large values of $p \sim 3$ have been also reported
in a GaAs/AlGaAs multi-quantum-well structure
\cite{Wernnberg_PRB1986} and a copper film\cite{Roukes_PRL1985}
where the electron-phonon interaction dominates the inelastic scattering time.
Our result $p$ = 3.3 $\pm$ 0.3 seems to be consistent
with these cases of  strong electron-phonon coupling and low mobility.
In MBE grown thin films of non-magnetic (Bi,Sb)$_2$Te$_3$ thin films, 
the $p$ value is reported as $0.5 < p < 1$
from the anti-weak localization analysis of magneto-conductivity\cite{JLiao_NCommun2016}.
This result implies that incorporation of the magnetic elements 
causes the enhancement of $p$.
Besides the phonon scatterings,
the effect of magnon scatterings\cite{KYasuda_PRL2016} should be taken
into consideration as a decoherence source of electrons.

To summarize, the magnetization rotation driven QPT between the QAH insulator and the
hybridization-induced trivial insulator is studied as a function of excitation current.
The sharpness of the QPT $d\sigma_{xy}/d\cos\theta$
is found to follow a power law current dependence.
Combined with the result of the temperature scaling,
the critical exponents for the localization length $\nu$ and for the coherence length $p$
are evaluated experimentally.
The obtained value of $\nu$ is consistent with the results of the Chalker-Coddington model,
pointing that the QAH insulator to trivial insulator transition  can be described by a quantum percolation of chiral edge channels.

We acknowledge valuable discussions with Akira Furusaki,  Naoto Nagaosa,  Kenji Yasuda, and Yayu Wang.
This study was supported by JSPS/MEXT Grant-in-Aid for Scientific Research
(No. 15H05853, 15H05867, 17J03179, 18H04229, and 18H01155), 
and CREST(JPMJCR16F1), JST.

\end{document}